\documentclass[pre,preprint,groupedaddress,showkeys,showpacs,floatfix]%
{revtex4}
\usepackage{graphicx,amsmath,amssymb,bm}
\usepackage{ifthen}
\graphicspath{{Figs/}}
\newboolean{doscol}\setboolean{doscol}{true}

\newlength{\figlength}
\newcommand{\ben}{\begin{displaymath}}
\newcommand{\een}{\end{displaymath}}
\newcommand{\be}{\begin{equation}}
\newcommand{\ee}{\end{equation}}
\newcommand{\bea}{\begin{eqnarray}}
\newcommand{\eea}{\end{eqnarray}}
\newcommand{\bean}{\begin{eqnarray*}}
\newcommand{\eean}{\end{eqnarray*}}

\newcommand{\pdone}[2]{{\partial#1 \over \partial#2}}

\newcommand{\dec}[2]{{#1 \times 10^{#2}}}

\newcommand{\pbpt}{{PB{\_}PT}}
\newcommand{\pbat}{{PB{\_}AT}}
\newcommand{\abpt}{{AB{\_}PT}}
\newcommand{\abat}{{AB{\_}AT}}

\begin{document}

\title{Mixing in  thermally stratified nonlinear spin-up with sources and sinks}

\author{Meline\, Baghdasarian}
\affiliation{Department of Mechanical Engineering,
California State University, Los Angeles, CA 90032, USA}
\author{J.\, Rafael Pacheco}\email{rpacheco@asu.edu}
\affiliation{SAP Americas Inc., Scottsdale AZ 85251, USA}
\affiliation{School of Mathematical and Statistical Sciences,
Arizona State University, Tempe, AZ 85287, USA}
\affiliation{Environmental Fluid Dynamics Laboratories,
 Department of Civil Engineering and Geological Sciences,
 The University of Notre Dame, South Bend, IN 46556, USA}
\author{Roberto Verzicco}
\affiliation{Dipartimento di Ingegneria Meccanica, Universita' di Roma 
``Tor Vergata'', Via del Politecnico 1, 00133, Roma, Italy}
\affiliation{PoF, University of Twente,  7500 AE Enschede, The Netherlands} 
\author{Arturo Pacheco-Vega}
\affiliation{Department of Mechanical Engineering,
California State University, Los Angeles, CA 90032, USA}

\begin{abstract}
Stratified spin-up experiments in enclosed cylinders have reported the
presence of small pockets of well-mixed fluids but quantitative measurements
of the mixedness of the fluid has been lacking. Previous numerical simulations
have not addressed these measurements.
Here we present numerical simulations that address how
the combined effect of spin-up and thermal boundary conditions
enhances or hinders mixing of a fluid in a cylinder. Measurements of
efficiency of mixing are based on the variance of temperature and explained
in terms of the potential energy available. The numerical simulations of the
Navier--Stokes equations for the problem with different sets of thermal boundary
conditions at the horizontal walls helped shed some light on the physical
mechanisms of mixing, for which a clear explanation was lacking.
\end{abstract}




\maketitle

\section{Introduction}\label{sec:intro}
Stratified spin-up flow is a classical fluid mechanics problem that has received
considerable attention in recent years. The transient flow is created when the fluid,
either at rest or in the state of solid body rotation, experiences an increase in the rotation rate
and results in the propagation of stresses into the interior.
The dynamics of spin-up/down is particularly relevant to large-scale geophysical flows, for example
in situations where wind stresses in the open ocean and coastal regions
generate ocean gyres and can result in baroclinic motions
that distort the temperature field, turbulent mixing and
redistribution of heat fluxes \cite[]{All73,LiHe84,MaHu85,McWil85,Mon86,Ols91,GMacR93,MoFl04}.

The study of mechanisms leading to efficient mixing  has long been appreciated in the context of
stratified shear flows \cite[]{PeCa03} and thermal convection \cite[]{Tur73,Tur86,SIT96,Max97,DPCC08,WiYo09}.
For example, shear can increase mixing at stratified interfaces by triggering Kelvin-Helmholtz  (K-H)
instabilities and can produce turbulence via interaction of Reynolds stresses \cite[]{Fer91,StFe01}.
Turbulence in the ocean can also be generated by another mechanisms, including
mean velocity shear, breaking of surface or internal waves and surface cooling.

Motions associated with upwelling are known to cause localized mixing  \citep{Mon86,WeHel04}.
Since most of the time new water masses are formed at the surface by cooling, and their
spin-up is clearly of utility in determining ensuing flow patters, it will be
helpful to understand how the spin of water masses in basins subjected to different
thermal boundary conditions affect the mixing.
Laboratory experiment of salt-stratified spin-up in a cylinder have shown 
qualitative measures of mixing \cite[]{Ped87,Gre80,FUB02,FBU04}, and 
recent three-dimensional simulations have demonstrated 
how different sets of 
thermal boundary conditions at the horizontal walls
(adiabatic or fixed temperatures)  affect the time of formation of columnar baroclinic vortices~\cite[]{SPV10}.
Nevertheless, quantitative measurements of mixing and 
the physical mechanisms controlling its efficiency in spin-up has remained relatively unexplored.

In this paper, we study the spin-up of a thermally stratified flow in a cylindrical container
in a numerical setting. 
In addition to the two sets of thermal boundary conditions
already considered in \cite{SPV10,PaVe12},
we include a combination of (\emph{i}) prescribed temperature
at the bottom wall and adiabatic at the top, and (\emph{ii}) prescribed temperature at
the top wall and adiabatic at the bottom. 
The quest here is for a quantitative measure of mixing
for a variety of thermal boundary conditions potentially relevant to ocean flows.
Our procedure for determining the quality of mixing is based on the variance of 
temperature \cite{Thi12,Pac08,PPC11}
and on the available potential energy \cite{WLRD95,WiYo09}.
Quantifying mixing in the initial-value decaying problem must be interpreted
very differently when sources and sinks are present. Common belief assumes that the best
stirring to create mixing is either turbulent or exhibits chaotic trajectories. However
this depends on the source-sink configuration, so a straight forward answer is not possible.
We will address these features in the next sections.


\begin{figure}\medskip
   \centering
  \begin{minipage}[t]{.99\textwidth}
      \begin{center}
      \includegraphics[width=.90\textwidth]{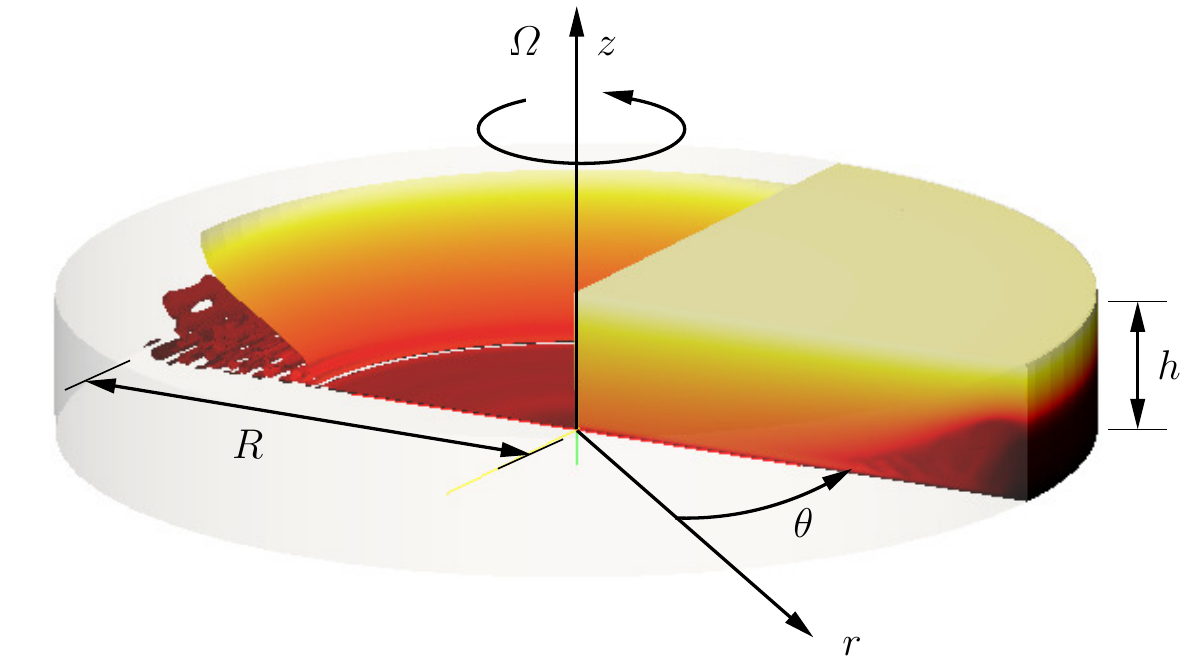}
      \end{center}
  \end{minipage}
        \caption{Schematic of the spin-up seven rotations after 
the cylinder is accelerated from the initial rotation rate
$\varOmega_i=\varOmega (1-\epsilon)$ to $\varOmega$.
The left and right quadrants show the vortex core
and the accumulation of cold fluid at the bottom corner respectively.}
\label{fig:geometry}
\end{figure}

\section{Governing equations and the numerical scheme}

Consider a Newtonian fluid of kinematic viscosity $\nu$, 
thermal diffusivity $\kappa$, and 
coefficient of volumetric expansion $\alpha$,
confined in a cylinder of radius radius $R$ and height $h$
where the gravity and rotation vectors are colinear,
as shown schematically in figure~\ref{fig:geometry}.
Initially, the fluid is thermally stratified in the vertical direction, with
a temperature difference of $\Delta T$ over $h$,
The flow is spun-up by the sudden change of background rotation
by the amount $\Delta \varOmega$ to a new rotation rate  $\varOmega$ from its
initial state $\varOmega_i = \varOmega(1-\epsilon)$ where $\epsilon = \Delta \varOmega / \varOmega$.
The system is non-dimensionalized using the flow depth $h$ as the
length scale, the inertial time $\varOmega^{-1}$ as the time scale and $\Delta T$ as the
temperature scale.  There are six non-dimensional parameters in this problem:
\begin{alignat*}{2}
 & \text{Aspect ratio:}    && \varGamma=R/h,\\
 & \text{Ekman number:} &\quad& E=\nu/ \varOmega h^2,\\
 & \text{Froude number:} && F = \varOmega^2 h/g, \\
 & \text{Burger number:} &&  B = N/ \varOmega, \\
 & \text{Prandtl number:} &&  Pr = \nu/\kappa, \\
 & \text{Rossby number:} && \epsilon = \Delta \varOmega/\varOmega,
\end{alignat*}
where $N=(\alpha g \Delta T/h)^{1/2}$ is the buoyancy frequency.
The non-dimensional governing equations are
\bea
(\partial_t  + {\bm u} \cdot \nabla) {\bm u} & = & \nabla p
+ B^2 \varTheta  {\bm e}_z
+ 2 {\bm u} \times  {\bm e}_z  - F \, B^2 \varTheta r {\bm e}_r +
E \nabla^2 {\bm u},  \label{eq:NS} \\
(\partial_t  + {\bm u} \cdot \nabla) \varTheta & = & Pr^{-1}E  \nabla^2 \varTheta, \qquad
\nabla \cdot {\bm u} =  0, 
\label{eq:PS}
\eea
where $\bm{u}$
is the velocity field in the rotating frame,
$(u,v,w)$ are the components of $\bm u$,  the cylindrical coordinates $(r,\theta,z)$
are the components of $\bm r$,
$p$ is the pressure (including gravitational and centrifugal contributions),
$\varTheta$ is the non-dimensional temperature.
The unit vectors in the radial and vertical directions are
$\bm{e}_r$ and $\bm{e}_z$  respectively.
The initial conditions in the rotating frame are $u=w=0, v=- \epsilon \, r$, and $\varTheta = z$,
the side-wall is no--slip and adiabatic, the top boundary is shear-free and the bottom wall non-slip,
and the lateral wall is insulated.
We shall focus our efforts on four sets of thermal boundary conditions applied to the
horizontal walls,  listed in table~\ref{table:cases}.
Two of these (\pbpt\, and \abat) were used in the analysis of \cite{PaVe12} to explain how
different sets of boundary conditions affect the time of formation of baroclinic vortices.

\begin{table}
\begin{center} {\footnotesize
\begin{tabular}{l|c|c} 
{Case}                                       & bottom wall ($z=0$) & top wall ($z=1$)  \\ \hline
\pbpt & $\varTheta=0$                     &  $\varTheta=1$                                    \\
\abpt & $\partial \varTheta/\partial z=0$ &  $\varTheta=1$                                    \\
\pbat & $\varTheta=0$                     &  $\partial \varTheta/\partial z=0$                \\
\abat & $\partial \varTheta/\partial z=0$ &  $\partial \varTheta/\partial z=0$                \\ 
\end{tabular} }
\end{center}
\caption {Boundary conditions used in the simulations.} 
\label{table:cases}
\end{table}

The governing equations \eqref{eq:NS}--\eqref{eq:PS} 
are discretized on a staggered
grid with the velocities at the faces and all the scalars in the
center of the computational cell; the resulting system of equations is
solved by a fractional-step method. 
The finite-difference solver is based on that described by
\cite{VeOr96} and has been tested in a wide variety of enclosed
cylindrical flows \cite[]{VeCa97,VLC97,SPV10,SPV10a,PLM11,PRZV11}, 
establishing resolution requirements over a wide range of parameters.

The grid is evenly spaced in the azimuthal direction while it
is non uniform in the radial and vertical direction in such a way to cluster
more computational points close to the solid (no--slip) boundaries where the largest
gradients occur. 
At least ten grid points were placed inside the bottom Ekman and side-wall boundary 
layers respectively, with $n_\theta  \times n_r \times n_z =  96 \times  351  \times 151$.
Details about the experimental and numerical test problems used for verification of the numerical code
and selection of number of grid points can be found in \citep{PaVe12}.


We split the variables into axisymmetric and nonaxisymmetric parts and employ the energy equation to quantify the azimuthal perturbations.
The axisymmetric part represents the mean flow (averaged quantities on the azimuth), while
the non-axisymmetric part corresponds to the flow perturbations.
For example, the velocity in  (\ref{eq:PS}) can be expressed as ${\bm u}(r,\theta,z) = \bar {\bm u}(r,z) + {\bm u}'(r,\theta,z)$,
where
\be
\bar {\bm u}(r,z) = \frac{1}{2 \pi} \int_0^{2 \pi} { \bm u}(r,\theta,z)  \,{\rm d }\theta. \label{eq:average}
\ee
Substituting (\ref{eq:average}) in the momentum equation (\ref{eq:PS}), taking the dot product with $\bm{u}'$, and
integrating over the entire domain $V$, yields the energy equation for the azimuthal disturbances
\bea
 \frac{{\rm d} e}{{\rm d}t} = \frac{\rm d}{{\rm d}t} \int_V \frac{1}{2}|\bm{u}'|^2 {\rm d}V & = &
- \int_V \bm{u}' \cdot (\bm{u}' \cdot \bm{\nabla} \bm{\bar u} ) {\rm d}V
- B^2 \int_V  \varTheta \ u_z' {\rm d}V   \nonumber \\
&&+ FB^2 \int_V  \varTheta \ r u_r' {\rm d}V
 - E   \int_V |\nabla \bm{u}'|^2 {\rm d}V = \sum_{i=1}^4 h_i.  \label{ekinetic}
\eea
The left-hand-side of (\ref{ekinetic}) represents the kinetic energy growth rate of the azimuthal disturbance due to
(${h_1}$) shear of the mean axisymmetric flow (barotropic production);
(${h_2}$) conversion of gravitational potential energy (baroclinic production);
(${h_3}$) conversion of centrifugal potential energy; and
(${h_4}$) viscous dissipation \cite[]{VLC97}.
A norm that is commonly used to quantify the mixing of the fluid is given
by the magnitude of the variance of the scalar $\varTheta$,
\be
\sigma = \frac{\langle \varTheta^2(r,\theta,z, t) \rangle - \langle \varTheta({r},\theta,z, t) \rangle^2}{\langle \varTheta^2({r},\theta,z, 0) \rangle - \langle \varTheta({r},\theta,z, 0) \rangle^2},
\label{norm}
\ee
where $\langle \cdot \rangle = 1/V \int_V \cdot {\rm d} V$.
In the presence of sources and sinks, the norm~(\ref{norm}) would reach
an asymptotic limit, and normalizing the global measure by the value it would
have in the absence of stirring, instead of the initial value, would be more helpful, i.e.
\be
\hat \sigma = \frac{\langle \varTheta^2({r},\theta,z, t) \rangle}{\langle \hat \varTheta^2({r},\theta,z, t) \rangle},
\label{hat_norm}
\ee
where $\hat \varTheta$ is the temperature due to diffusion only \cite{DoTh06}. 
Efficient mixing implies $ \hat \sigma < 1$ if the stirring decreases the
variance relative to molecular diffusion alone, which is not always the case
when sources or sinks are present.
We will describe the time-evolution of the solutions in terms of
the number of rotation $\tau (= t/2 \pi)$ instead of the normalized time $t$.

We also quantify the available potential energy for mixing ($PE_A$) by computing the difference between the total potential energy ($PE_T$) and the potential
energy of a reference state ($PE_R$) \cite{TsFe01,DPCC08,BWLS13}, that is the minimum potential energy that can be obtained through an adiabatic redistribution of temperature (density), 
\be
PE_A = PE_T - PE_R = \int_V (1 - z) \, \varTheta {\rm d}  V - \int_V  (1-z_R) {\varTheta}  \, {\rm d}  V.
\ee
Here, $z_R(\varTheta,\tau) $ is the vertical coordinate of the reference state  (where all the temperature surfaces are horizontal). 
The vertical height of the reference state $z_R$ can be computed
in different manners, for example, by reorganizing the vertical position of layers in the reference state according to their density 
with the Heaviside step function $H$,
\be
z_R({r},\theta,z,\tau) = \frac{1}{\pi \varGamma^2}  \int_{V} H[\varTheta({r},\theta,z,\tau) - \varTheta({r}', \theta', z', \tau)]  \, {\rm d}  V',
\label{eq:ZHeavyside}
\ee
or by computing the probability density function $\lambda(\varTheta)$ of the temperature,
\be
\lambda(\tilde \varTheta) = \frac{1}{V} \int_V \delta (\tilde \varTheta - \varTheta) \, {\rm d}  V.
\label{eq:PDF}
\ee
We evaluated numerically the  probability density function $\lambda(\varTheta)$ by 
scanning the temperature field and placing its values into a bin and
by normalizing the number of control volumes in each bin.
The reference position  $z_R(\varTheta)$ is obtained  using the probability density function from  (\ref{eq:PDF}) from
\be
z_R(\varTheta) = 1 -  \int_{\varTheta}^{\varTheta_M}  \lambda(\tilde \varTheta)\,  {\rm d}  {\tilde \varTheta},
\label{eq:newZ}
\ee
where the nondimensional height of the domain is 1 and $\varTheta_M$ is the maximum value of the temperature at time $\tau$. The potential
energy of the reference state $PE_R$ can now be obtained from 
\be
PE_R = \pi \varGamma^2 \int_{0}^{1} (1 - z_R) \, \varTheta\,   {\rm d}  z_R.
\label{eq:PER}
\ee

The parametric studies of \cite{FBU04,PaVe12} suggest that for 
$\varGamma < 1$ the spin-up is less prone to become non-axisymmetric, 
therefore we restricted the values of the Rossby numbers
to $\epsilon \in [0.5,1]$, and fixed the aspect ratio at $\varGamma = 3.3$,
the Ekman number at $E=\dec{7.2}{-4}  $, the Froude number at $F=\dec{9.0}{-4} $, 
the Burger number at $B=2.52$ and the Prandtl number at $Pr = 6.85$.

\section{Results and discussion}\label{sec:results}

Before discussing how  different boundary conditions affect the mixing, it is useful to briefly
review the flow dynamics addressed in \cite{SPV10,PaVe12}. 
Spin-up is a typical example of baroclinicity
whose dynamics is dictated by the equation for absolute vorticity $\bm{\omega}$.
Taking the curl  of (\ref{eq:NS}) yields 
\be
(\partial_t  + {\bm u} \cdot \nabla) {\bm \omega} =  {\bm \omega} \cdot \nabla  {\bm u}
+ B^2    \nabla \varTheta  \times {\bm e}_z
+ E \, \nabla^2 {\bm \omega}.  \label{eq:vorticity}
\ee
The first term on the right-hand side is responsible for vortex stretching and tilting, the
second accounts for baroclinic vorticity and the third term
represents vorticity diffusion.
The production of barotropic vorticity can be expressed in its components as
\be
\bm{\omega}_b = B^2 \nabla \varTheta  \times {\bm e}_z =B^2 \left(\frac{1}{r} \pdone{}{\theta} \bm{e}_r
                                   - \pdone{}{r} \bm{e}_\theta \right) \varTheta. \label{eq:baroclinic}
\ee
When the motion of the flow is initiated by the sudden increase in rotation rate,
Ekman transport along the bottom boundary layer pushes fluid radially outwards and
forms well-mixed corner regions that rotate faster than the interior.
The stable stratification causes the azimuthal flow to develop
vertical shear owing to the strong deformation of the isotherms
developing an unstable system that can convert
potential energy into kinetic energy.
The kinetic energy dissipates through friction and reduces 
the temperature contrast through temperature advection.
This is a common feature
of a stratified spin-up flow regardless of the thermal boundary 
conditions imposed on the horizontal walls. 

\begin{figure}
\begin{center}
\begin{tabular}{cc}
\multicolumn{2}{c}{$(a)$ $\tau=0$}  \\
\includegraphics[width=0.48\linewidth]{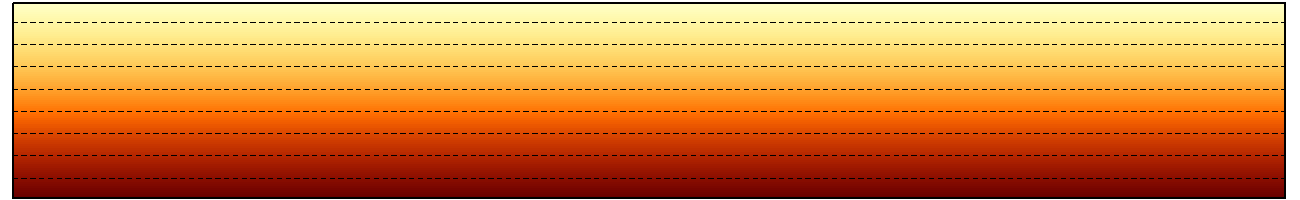}&
\includegraphics[width=0.48\linewidth]{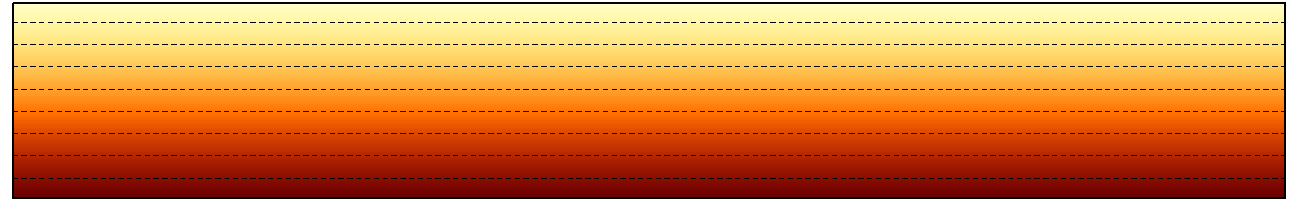}\\
\multicolumn{2}{c}{$(b)$ $\tau=10$}  \\
\includegraphics[width=0.48\linewidth]{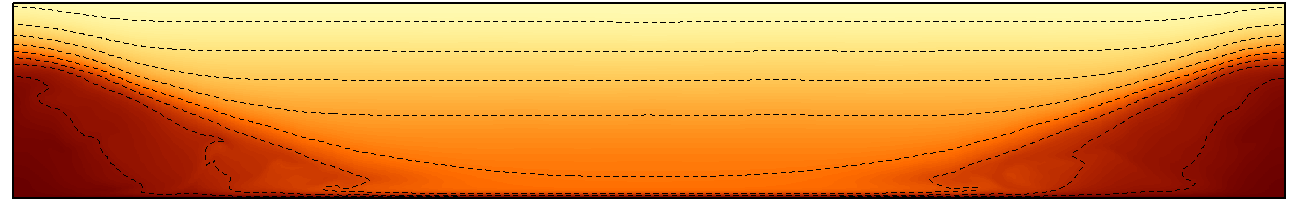}&
\includegraphics[width=0.48\linewidth]{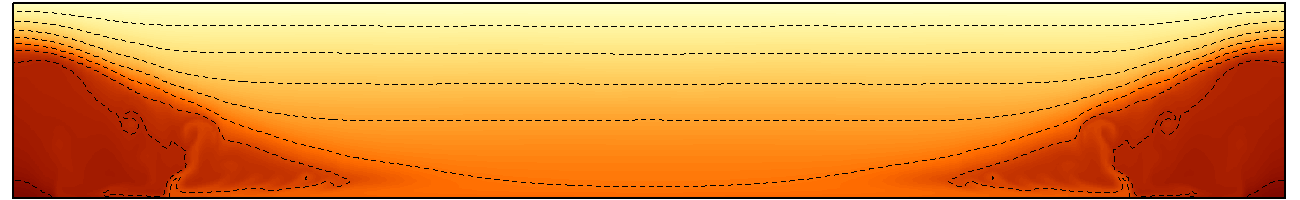}\\
\multicolumn{2}{c}{$(c)$ $\tau=30$}  \\
\includegraphics[width=0.48\linewidth]{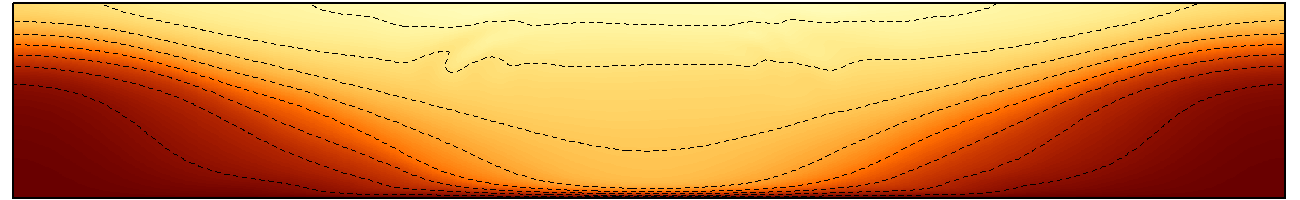}&
\includegraphics[width=0.48\linewidth]{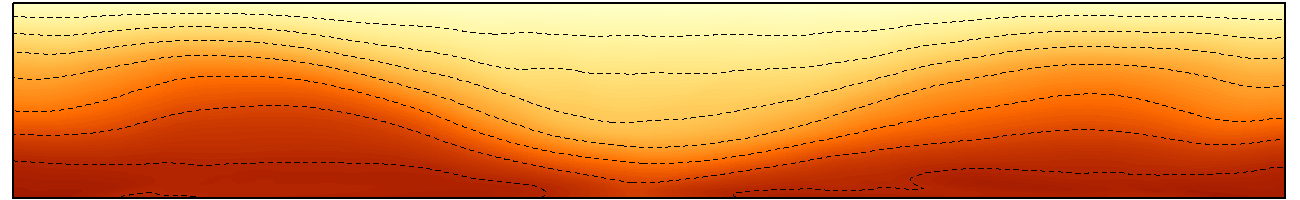}\\
\multicolumn{2}{c}{$(d)$ $\tau=50$}  \\
\includegraphics[width=0.48\linewidth]{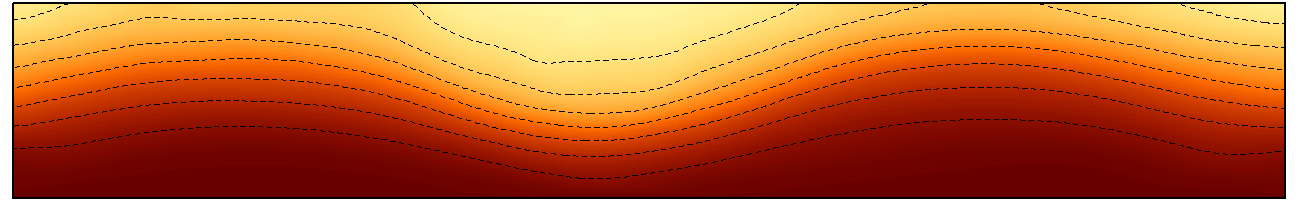}&
\includegraphics[width=0.48\linewidth]{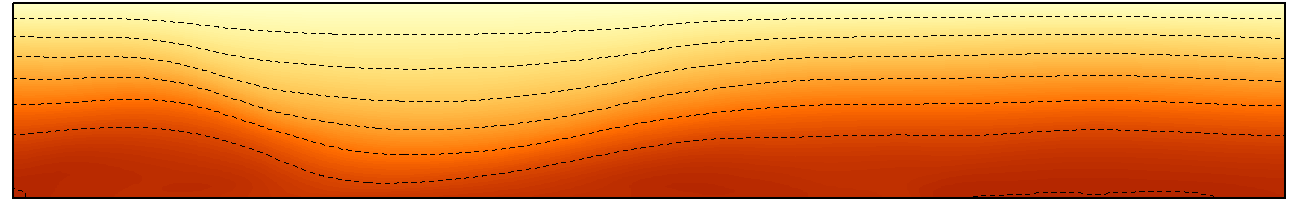}\\
\multicolumn{2}{c}{$(e)$ $\tau=70$}  \\
\includegraphics[width=0.48\linewidth]{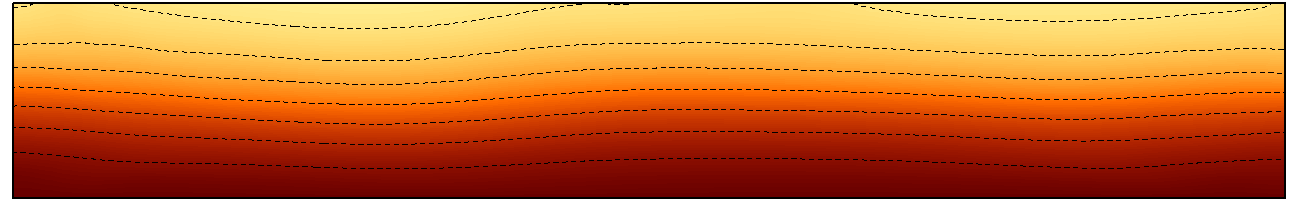}&
\includegraphics[width=0.48\linewidth]{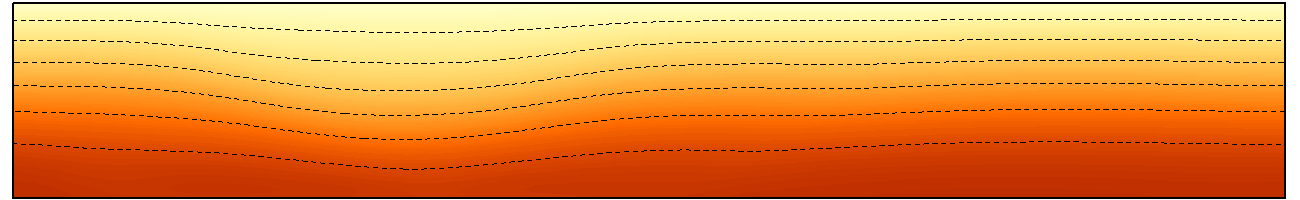}\\
\multicolumn{2}{c}{$(f)$ $\tau=90$}  \\
\includegraphics[width=0.48\linewidth]{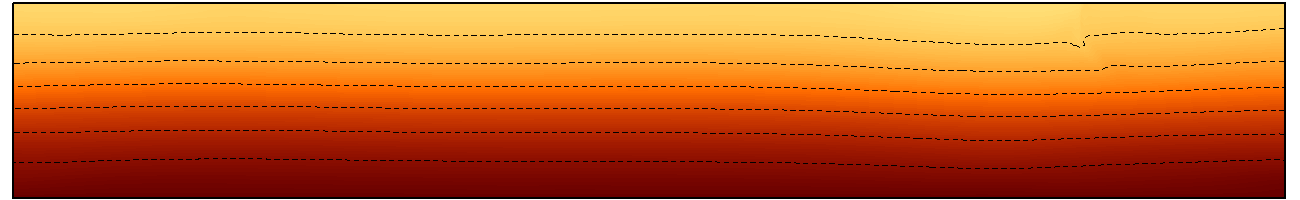}&
\includegraphics[width=0.48\linewidth]{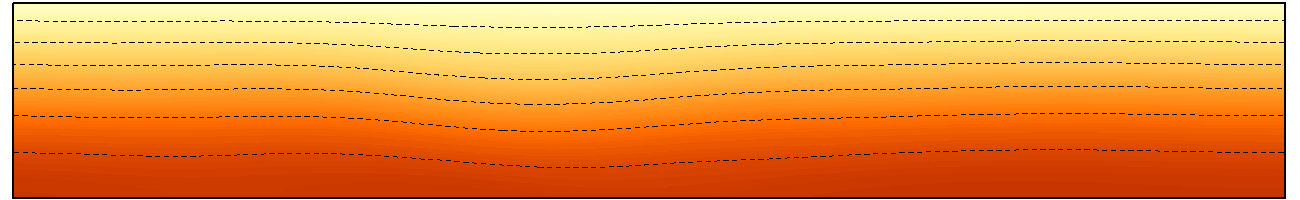}\\
\multicolumn{2}{c}{$(g)$ $\tau=110$}  \\
\includegraphics[width=0.48\linewidth]{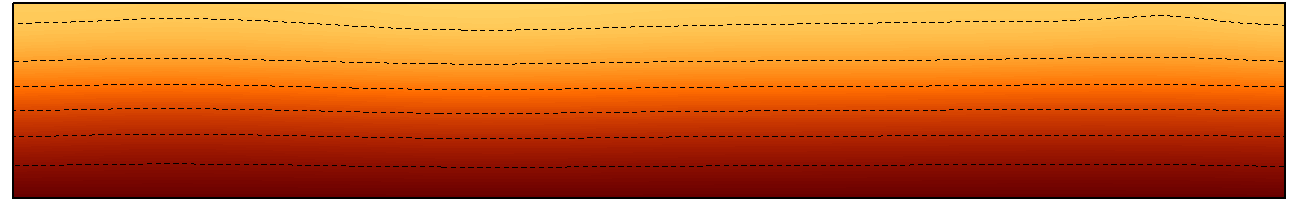}&
\includegraphics[width=0.48\linewidth]{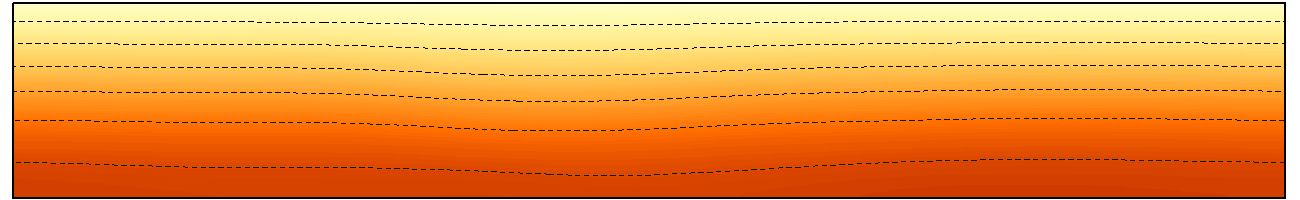}\\
\multicolumn{2}{c}{$(h)$ $\tau=130$}  \\
\includegraphics[width=0.48\linewidth]{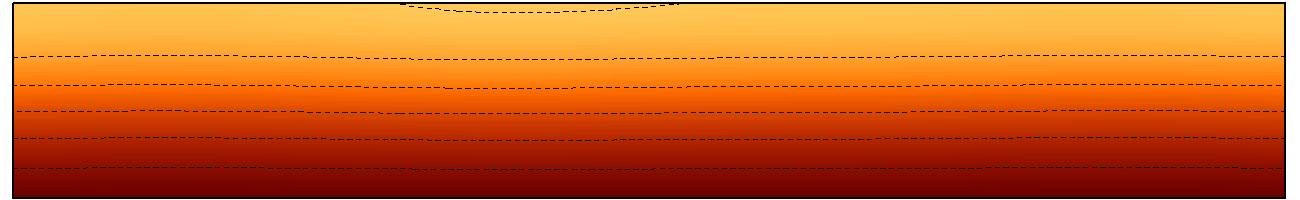}&
\includegraphics[width=0.48\linewidth]{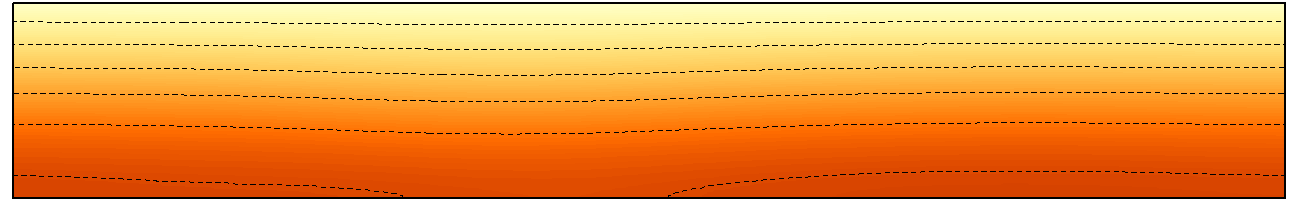}\\
Case: \pbat & Case: \abpt  \\
\end{tabular}
\end{center}
     \caption{(Color online) Contours of temperature $\varTheta$ on the planes $\theta=0-\pi$
              at $\epsilon = 1$.
              At $\tau=0$ there are 10 linearly spaced contour-levels in the
              range $\varTheta=[0,1]$.  The figures in the left column correspond to \pbat, and those in the right column
              to \abpt. See the supplementary movies for animations.}
\label{fig:temperaturee100}
\end{figure}

The numerical simulations of \cite{SPV10} for nonlinear spin-up and
large aspect ratios demonstrated that after the initial phase of motion,
the resulting stratification originating from different boundary conditions 
triggered different instabilities.
When the cylinder walls were adiabatic (\abat), 
the vortex-core became baroclinically unstable,
breaking up into different lenses.
For isothermal boundary conditions (\pbpt)
the baroclinic perturbation decayed, the vortex-core
began to oscillate and after several tens of rotations the flow 
broke into several columnar vortex structures as the flow
returned to a state of linear stratification.

How and if the unstable system develops columnar vortices
was addressed in the parametric study of \cite{PaVe12}. They explained
that when the temperatures are prescribed, the flow becomes three-dimensional
due to small azimuthal variations of temperature leading to an
increase in the baroclinic vorticity through (\ref{eq:baroclinic}). The
baroclinic vorticity in the radial component
$r^{-1}\partial \varTheta/\partial \theta$
excites the kinetic energies in the asymmetric azimuthal Fourier components,
and compensates for the decrease in $\partial \varTheta / \partial r$.
The growth of baroclinic vorticity produced by the azimuthal variations of
temperature enhances the axial vorticity, and this in turn
increases the vertical shear around the boundary of the central
vortex-core, further deforming the
isotherms on the $z$-plane.
This local advective heat transport enhances azimuthal temperature gradients
completing the feedback cycle causing the core vortex to break.

\begin{figure}
\begin{center}
\begin{tabular}{cc}
\multicolumn{2}{c}{$(a)$ $r=0.5$}  \\
\includegraphics[width=0.49\linewidth]{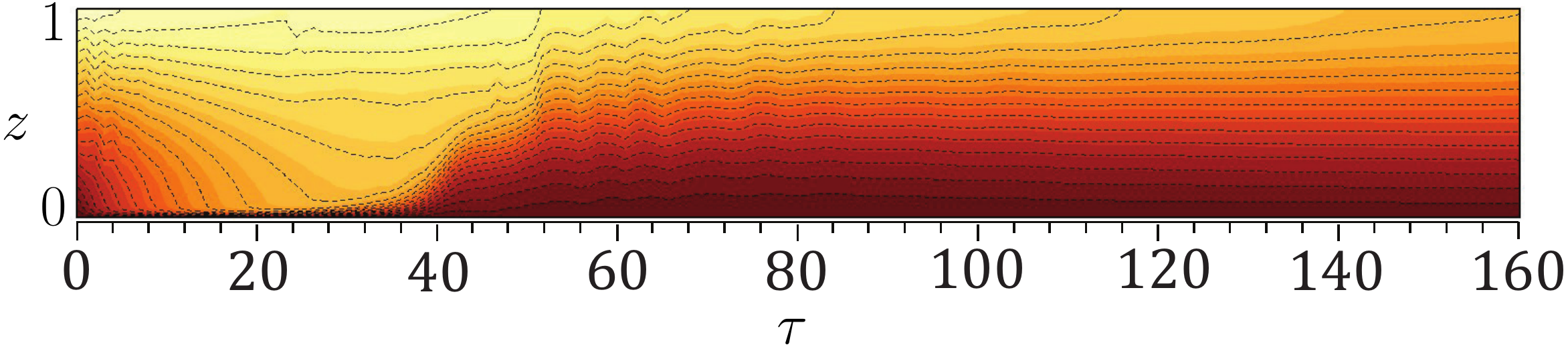}&
\includegraphics[width=0.49\linewidth]{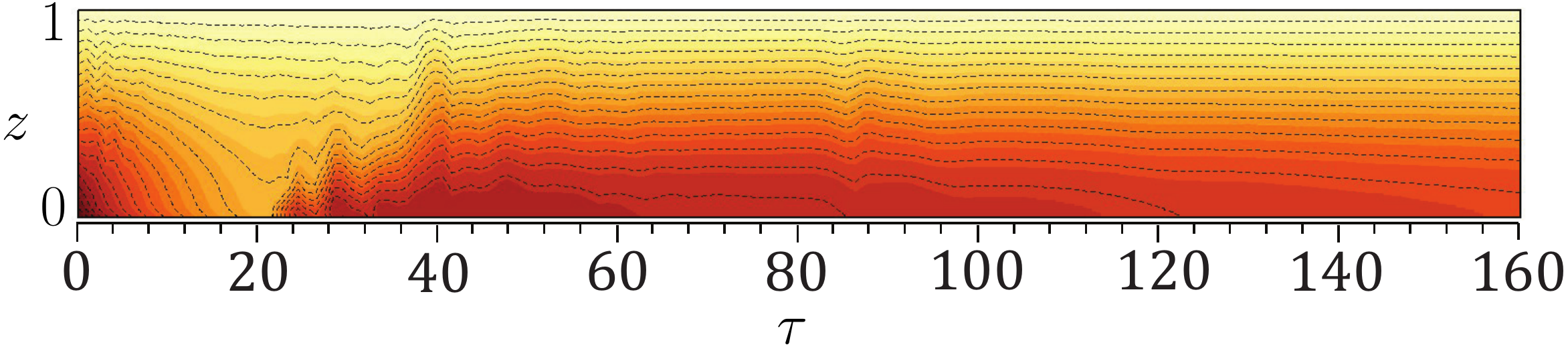}\\
\multicolumn{2}{c}{$(b)$ $r=1.7$}  \\
\includegraphics[width=0.49\linewidth]{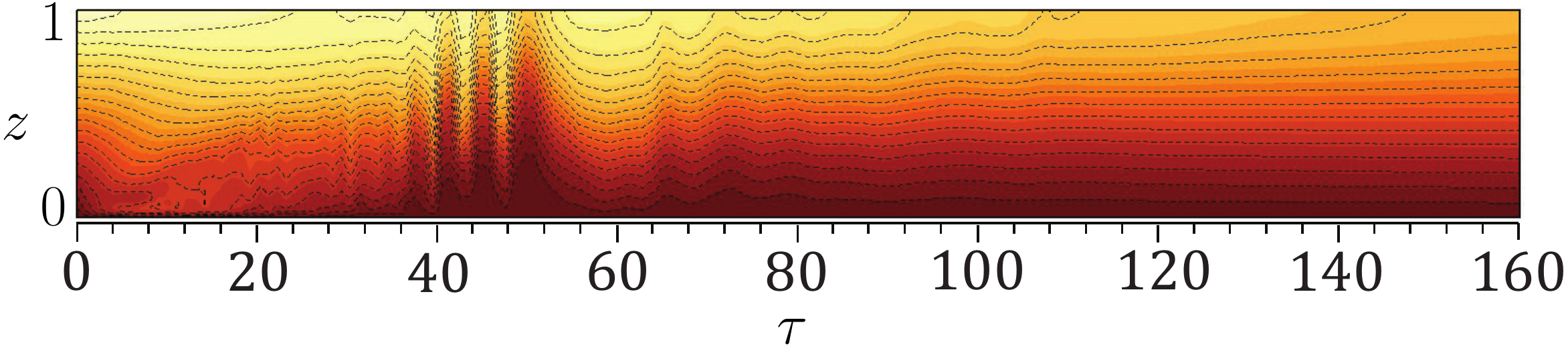}&
\includegraphics[width=0.49\linewidth]{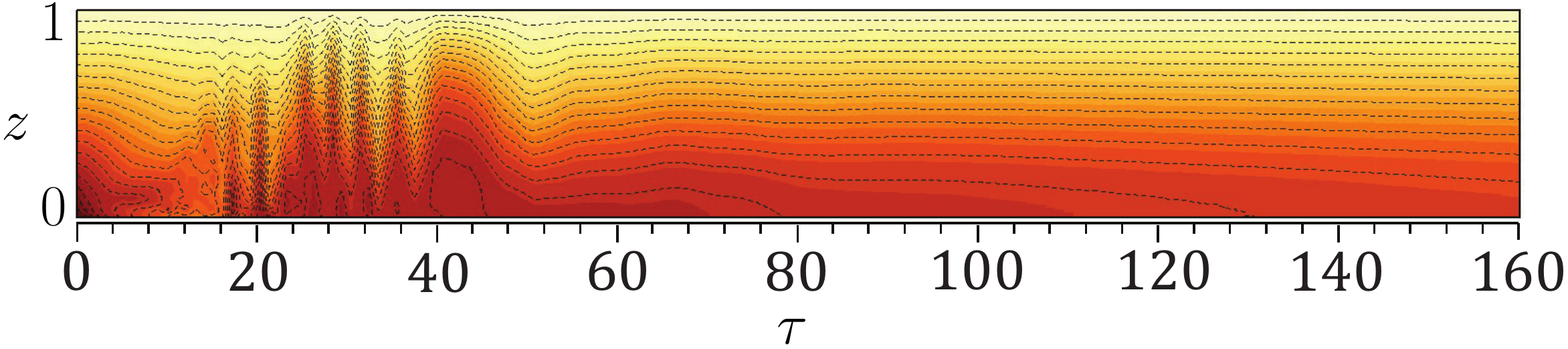}\\
\multicolumn{2}{c}{$(c)$ $r=3.2$}  \\
\includegraphics[width=0.49\linewidth]{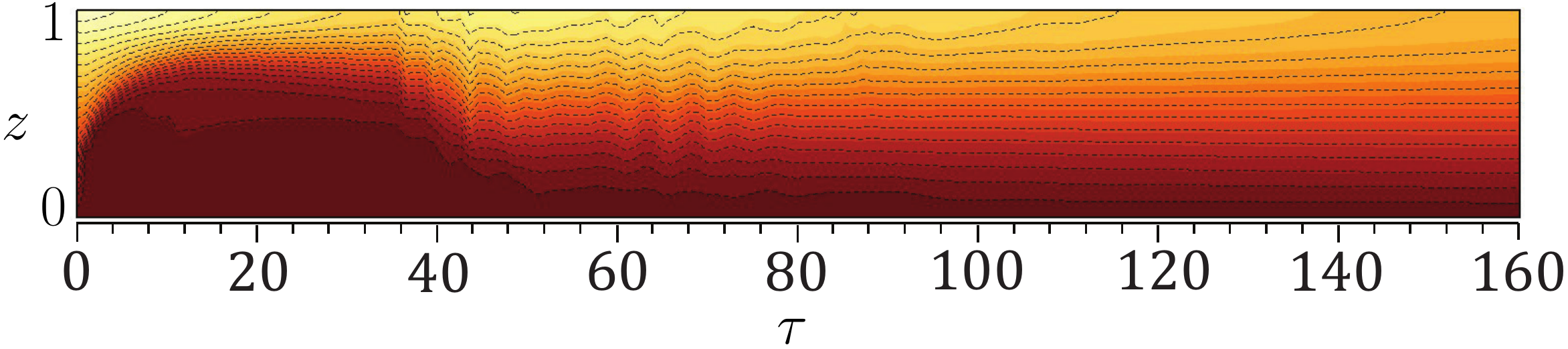}&
\includegraphics[width=0.49\linewidth]{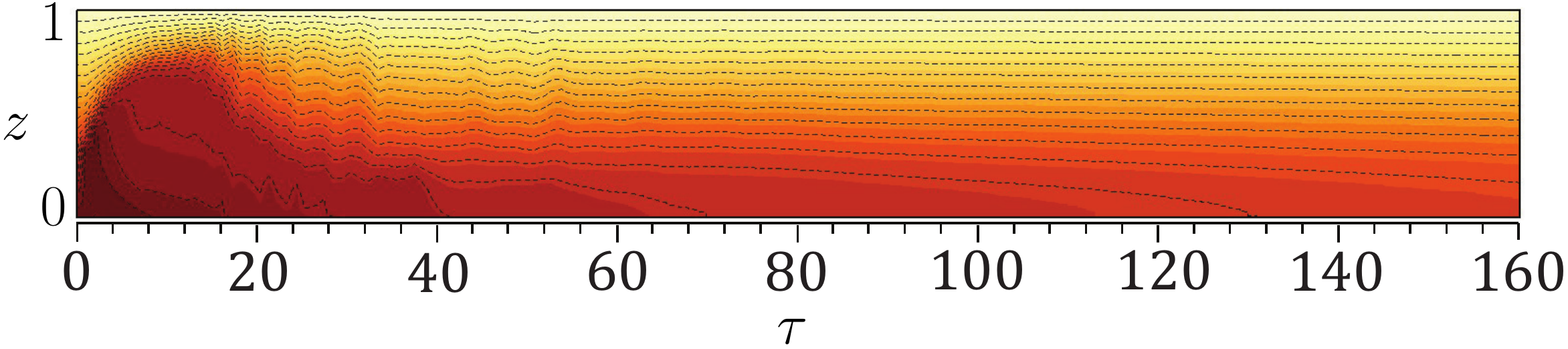}\\
Case: \pbat & Case: \abpt  \\
\end{tabular}
\end{center}
     \caption{(Color online) Spatio-temporal evolution of the temperature
              along a vertical line at $\theta=0$ and $r$ as indicated.
              The horizontal axis indicates time in the range $ 0 \le \tau \le 160$ and the vertical axis
              the location of the probes in the range $ 0 \le z \le 1$.
              At $\tau=0$ there are 10 linearly spaced contour-levels in the range $\varTheta=[0,1]$.
              The figures in the left column correspond to \pbat, and those in the right column to \abpt.}
\label{fig:transient_all_epsilon}
\end{figure}

\begin{figure}
\begin{center}
\begin{tabular}{cc}
%
\multicolumn{2}{c}{$(a)$ $\epsilon=0.73$ }  \\
\includegraphics[width=0.45\linewidth]{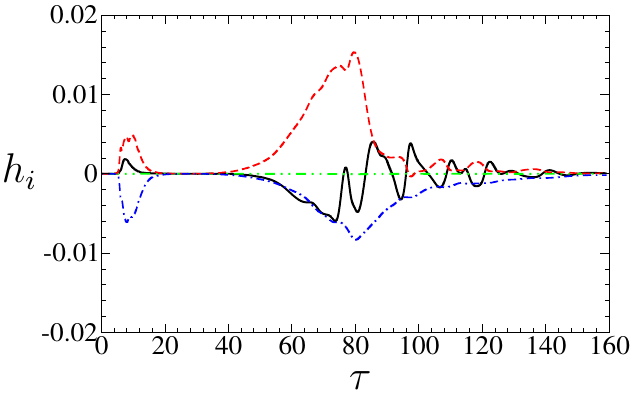}&
\includegraphics[width=0.45\linewidth]{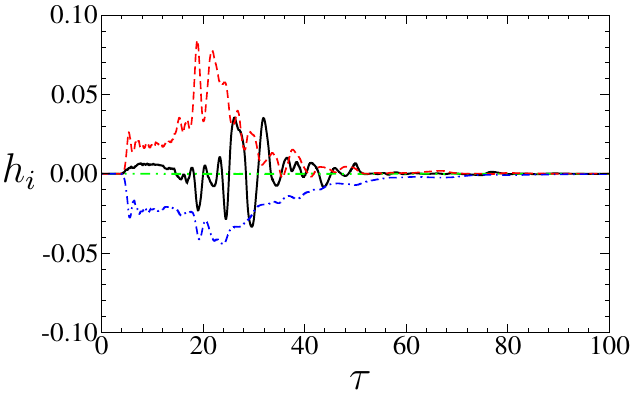}\\[6pt]
\multicolumn{2}{c}{$(b)$ $\epsilon=1$ }  \\
\includegraphics[width=0.45\linewidth]{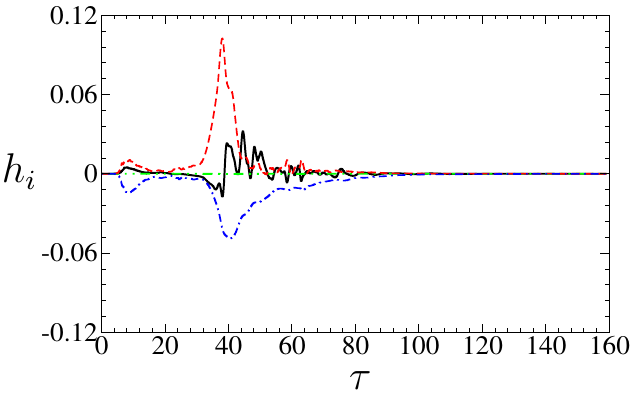}&
\includegraphics[width=0.45\linewidth]{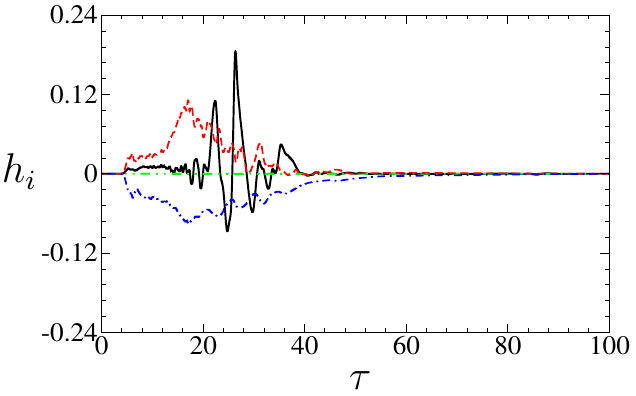}\\[6pt]
Case: \pbat & Case: \abpt  \\
\end{tabular}
\end{center}
     \caption{(Color online) Time evolution of $h_i$-terms in the rate of change of kinetic
              energy of azimuthal perturbations at $\epsilon$ as indicated.
              The figures in the left correspond to \pbat, and those in the right to \abpt.
              Barotropic term $h_1$ (----- black);
              baroclinic term $h_2$ (--~-- red);
              centrifugal term $h_3$ (--$\cdot \cdot$-- green);
              viscous dissipation term $h_4$ (--~$\cdot$~-- blue).
              }
\label{fig:henergy}
\end{figure}


The typical evolution of the flow for \pbat\, and \abpt\, at $\epsilon = 1$ is shown in 
figures~\ref{fig:temperaturee100} and \ref{fig:transient_all_epsilon}.
The sequence of images in figure~\ref{fig:temperaturee100} demonstrates the 
flow change on the planes $\theta = 0-\pi$.
At ten rotations ($\tau=10$), pockets of well-mixed cold fluid 
accumulate at the bottom corner separated from the 
core by a vortex. The left quadrant of figure~\ref{fig:geometry} 
illustrates the vortex core (front) using the $Q$-criterion.
At 30 rotations, the flow \abpt\, is three-dimensional with evidence
of internal waves. For \pbat\, and the same number of rotations,
the accumulation of well-mixed fluid at
the corner regions is still visible. The flush back of 
cold fluid, after the Ekman transport shuts down, occurs about 20 
rotations later for \pbat\, than for \abpt. The delay is influenced
mainly by the boundary condition at the bottom wall.
This event is clearly seen from the spatio-temporal evolution of 
temperature along a vertical line at three fixed radii of figure~\ref{fig:transient_all_epsilon}.
From this figure, we can also see the quality of mixing in the 
interior ($r=0.5$), around the interface of the core 
vortex during upwelling  ($r=1.7$), and near the lateral wall ($r=3.2$).
At early times the isotherms are compressed near the bottom wall ($r=0.5$) 
and near the top wall ($r=3.2$) as the cold fluid moves 
through the Ekman layer pushing cold fluid to the corner. 
When the Ekman pumping ceases, the secondary circulation reverses direction
and the cold fluid from the corner regions moves back to 
replace the warm fluid in the core. Near the adiabatic walls, the fluid
that is replaced is nearly homogeneous, whereas near the wall with 
prescribed temperatures the fluid remains stratified 
(figure~\ref{fig:transient_all_epsilon}).
The appearance of baroclinic waves is shown in figure~\ref{fig:transient_all_epsilon}($b$).
Notice that the baroclinic instability propagates from the vortex core to
both, the interior and to the outer wall.

The flow behavior for $\epsilon = \{0.5,0.73\}$ 
is similar to that of $\epsilon = 1$. The main difference is the time at which
the flow becomes three-dimensional, with the transition occurring later for
smaller Rossby numbers. This is better appreciated from the history
profile of azimuthal disturbances $h_i$. For comparison, we only
show values for $\epsilon=\{0.73,1\}$ in figure~\ref{fig:henergy}.
The left-hand-side of the figure shows 
the azimuthal disturbances for \pbat, and this demonstrates
a remarkable similarity with \pbpt, with
two distinct states in the flow development. The first is an increase in the energy of 
perturbation followed by a decay due to the horizontal realignment of the isotherms in the 
$\theta$-plane as the Ekman transport shuts down. 
The second is  characterized by an increase in the energy of perturbation
due to the baroclinic vorticity contribution in the radial component, which
compensates for the decrease in $\partial \varTheta / \partial r$ as 
explained by \cite{PaVe12}. 

The path to three-dimensionality of the flow for \abpt\, is very similar to \abat\, as well, with
the baroclinic disturbance remaining positive
until it reaches a global maximum. The barotropic term initially contributes
to the instability, and then extracts energy from the mean flow. The viscous dissipation
as expected is negative and the centrifugal term negligible.
\begin{figure}
\begin{center}
\begin{tabular}{cc}
\multicolumn{2}{c}{$(a)$  }  \\
\includegraphics[width=0.49\linewidth]{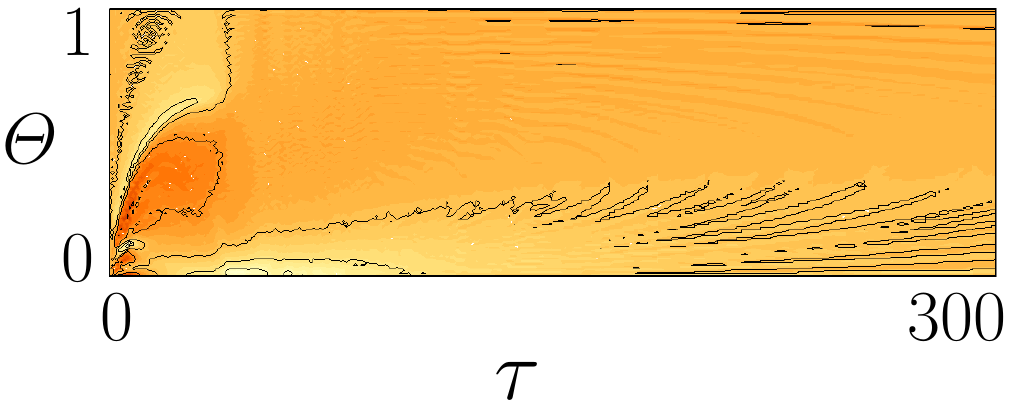}&
\includegraphics[width=0.49\linewidth]{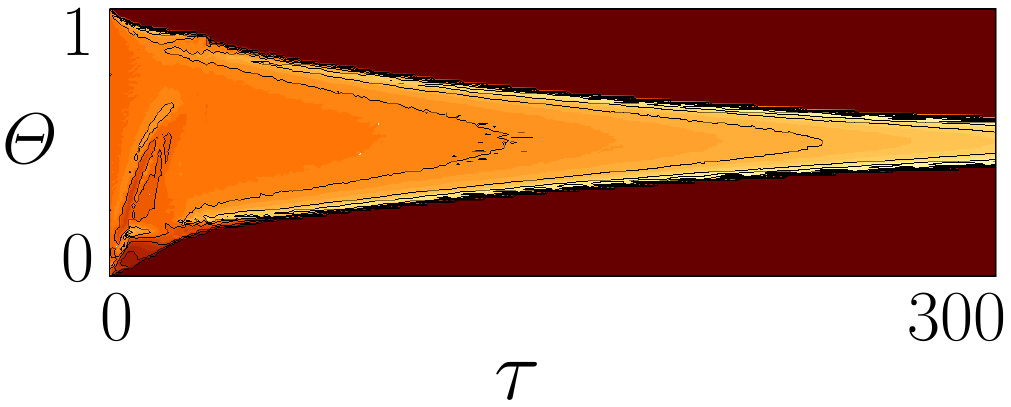}\\[2pt]
\multicolumn{2}{c}{$(b)$  }  \\
\includegraphics[width=0.49\linewidth]{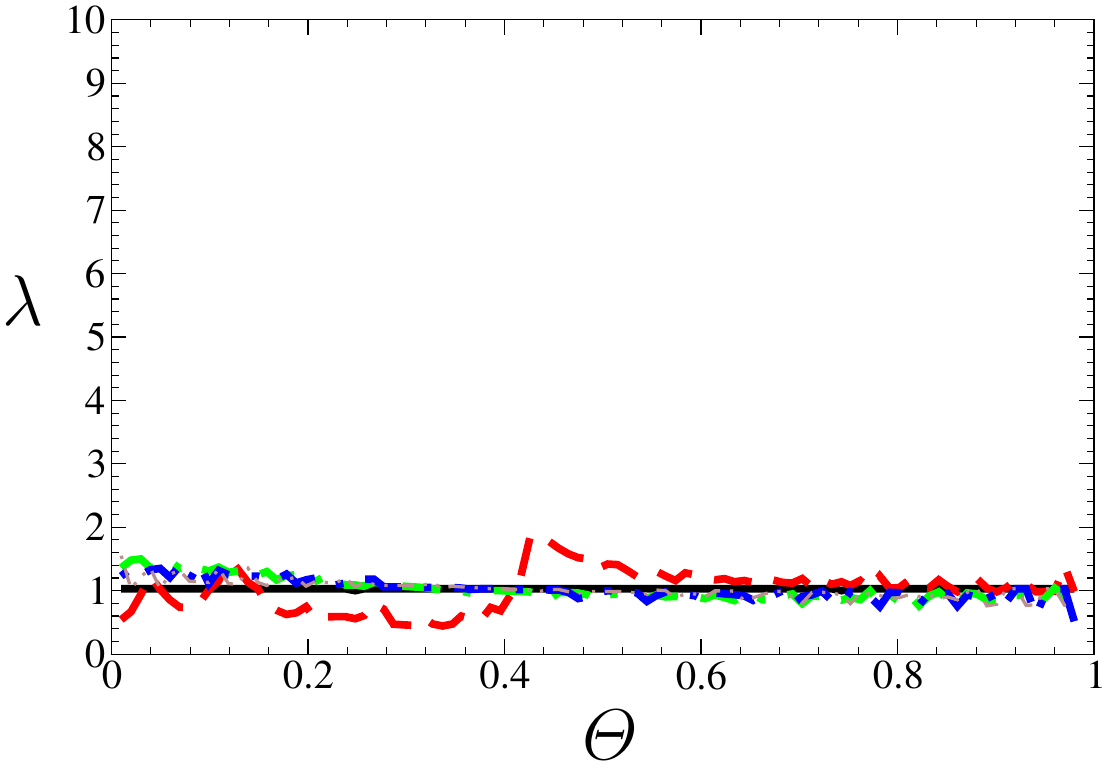}&
\includegraphics[width=0.49\linewidth]{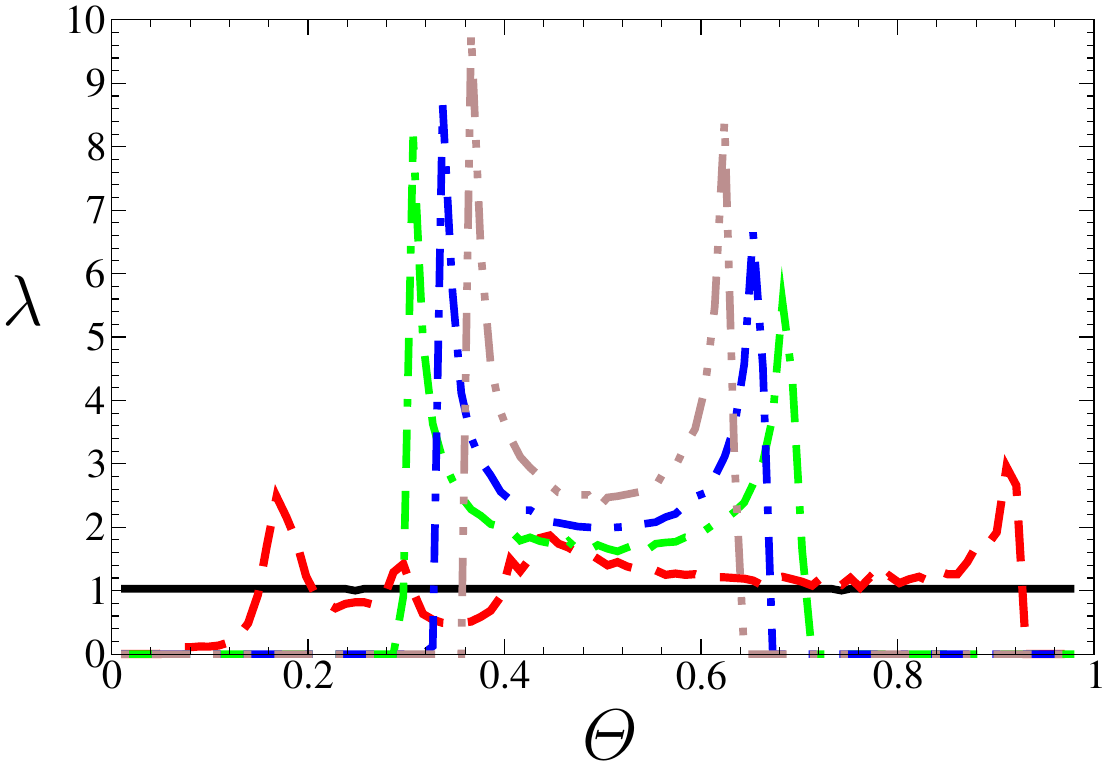}\\[2pt]
%
Case: \pbpt & Case: \abat  \\
\end{tabular}
\end{center}
     \caption{(Color online) $(a)$ Contours of probability density $\lambda(\varTheta)$ 
              as function of number of rotations $\tau$ and $(b)$ cross-sections
              of $\lambda$ at: ----- (black) $\tau=0$; -- -- (red) $\tau=10$; 
              -- -- $\cdot$ -- -- (green) $\tau=125$; -- $\cdot$ -- (blue) $\tau=160$;
              -- $\cdot \cdot $  -- (brown) $\tau=200$. The Rossby number is $\epsilon = 1$.
              }
\label{fig:pdf_evolution_IB}
\end{figure}
\begin{figure}
\begin{center}
\begin{tabular}{cc}
\multicolumn{2}{c}{$(a)$  }  \\
\includegraphics[width=0.49\linewidth]{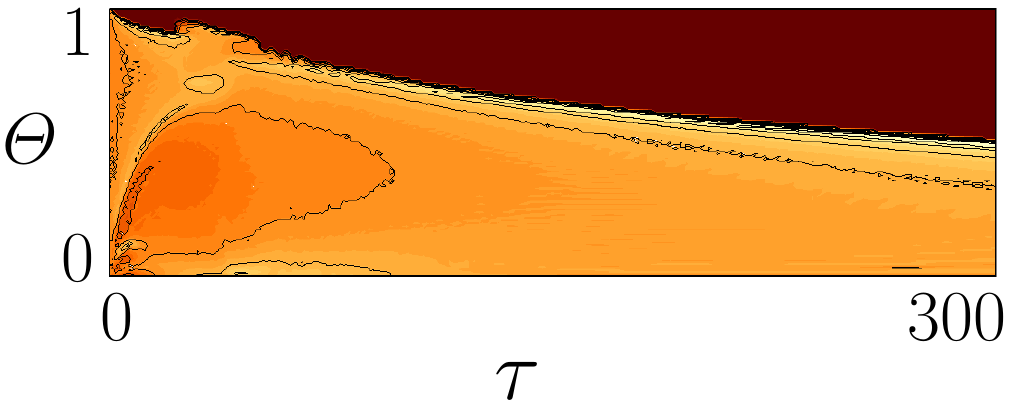}&
\includegraphics[width=0.49\linewidth]{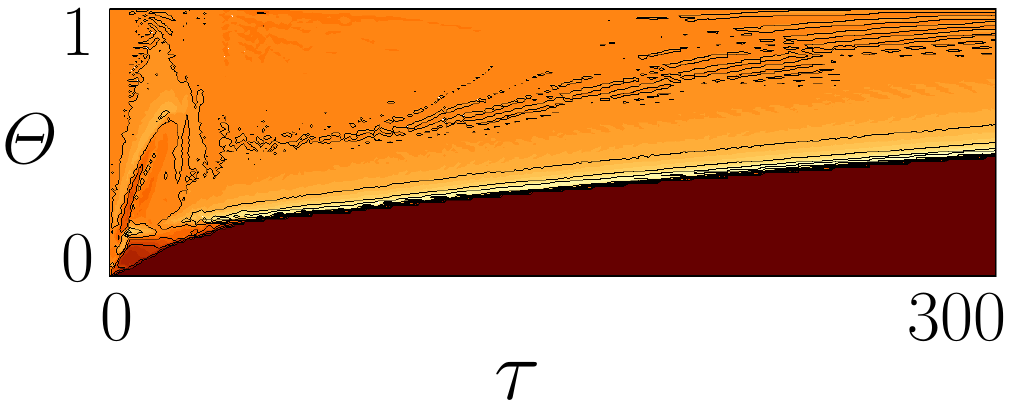}\\[2pt]
\multicolumn{2}{c}{$(b)$  }  \\
\includegraphics[width=0.49\linewidth]{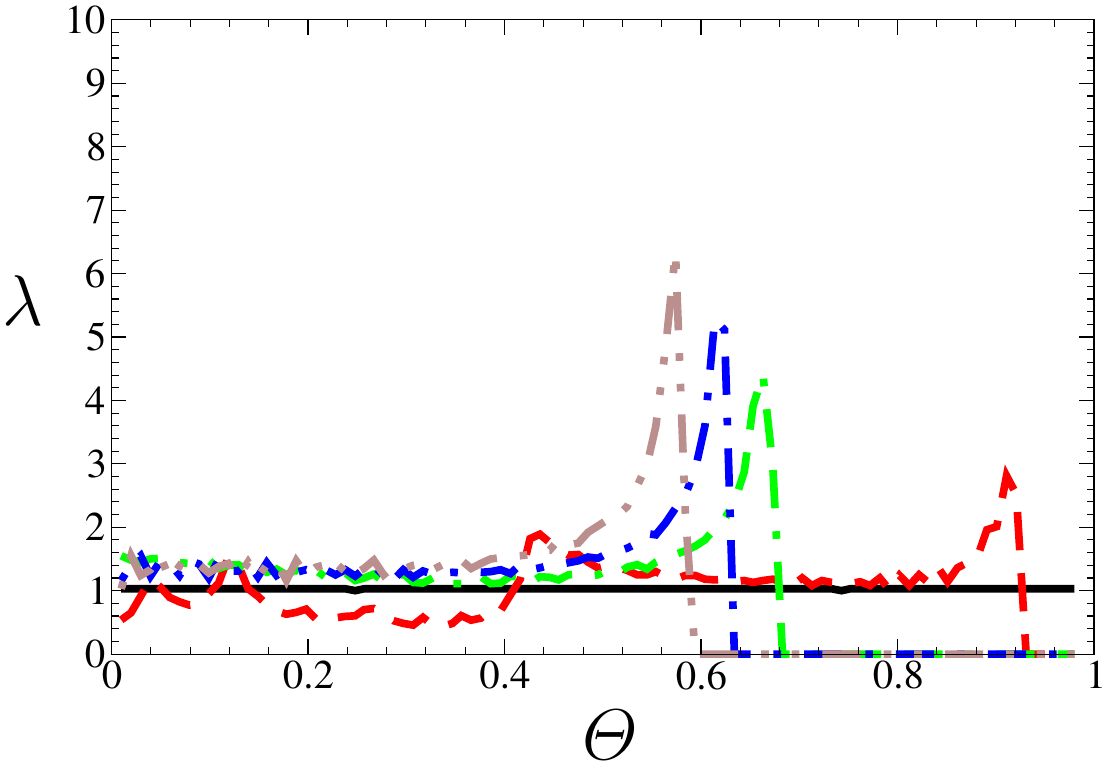}&
\includegraphics[width=0.49\linewidth]{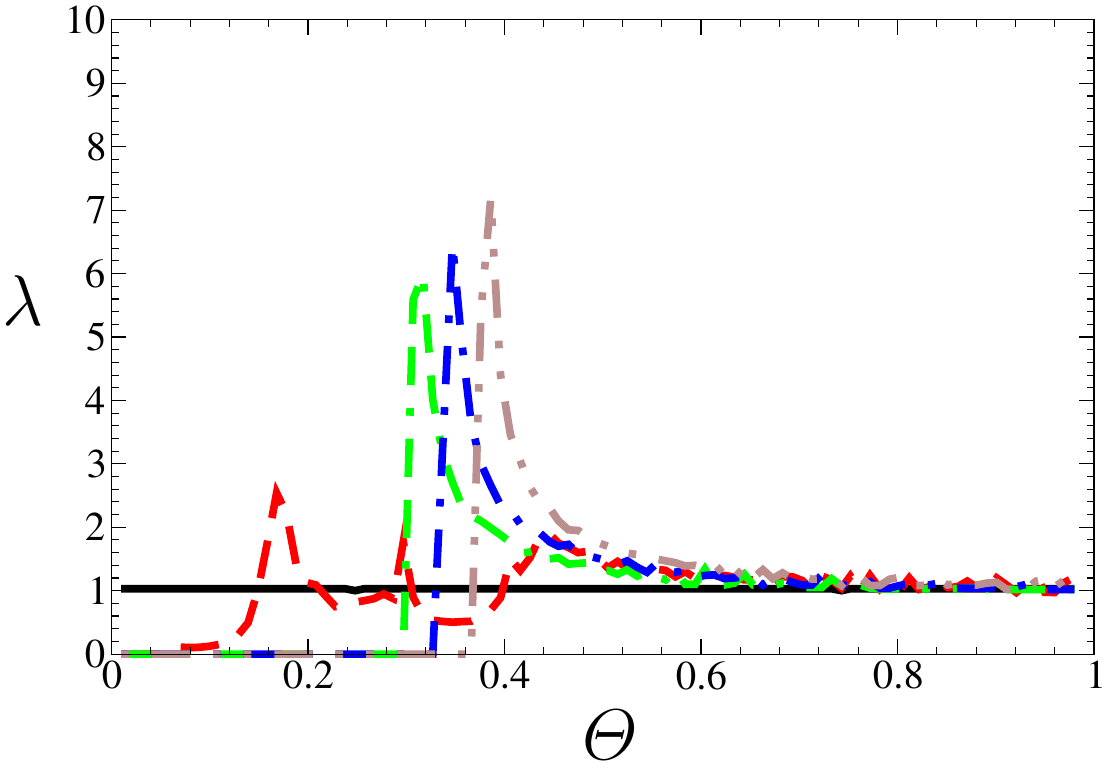}\\[2pt]
%
Case: \pbat & Case: \abpt  \\
\end{tabular}
\end{center}
     \caption{(Color online) $(a)$ Contours of probability density $\lambda(\varTheta)$ 
              as function of number of rotations $\tau$ and $(b)$ cross-sections
              of $\lambda$ at: ----- (black) $\tau=0$; -- -- (red) $\tau=10$; 
              -- -- $\cdot$ -- -- (green) $\tau=125$; -- $\cdot$ -- (blue) $\tau=160$;
              -- $\cdot \cdot $  -- (brown) $\tau=200$. The Rossby number is $\epsilon = 1$.
              }
\label{fig:pdf_evolution_IB_e100}
\end{figure}

The probability density function $\lambda(\varTheta)$ is a good indicator
of how the temperature is spatially distributed during spin-up. This is
evaluated numerically by scanning the temperature field, placing its values into 
bins and normalizing the values by the number of control volumes in each bin.
Contours of $\lambda(\varTheta)$ and cross-sections at various numbers of rotation
are shown in figures~\ref{fig:pdf_evolution_IB}--\ref{fig:pdf_evolution_IB_e100}
at $\epsilon = 1$ to show the spatio-temporal distribution of temperature
for the different sets of boundary conditions considered in this study.
At late times, the linear stratification for \pbpt\, is almost recovered, whereas
for \abat, the distribution of temperature is bi-modal, with the asymptotic values of
temperature concentrating around the mean $\langle\varTheta\rangle=0.5$.
The time evolution of $\lambda(\varTheta)$ for \pbat\, and \abpt\, at $\epsilon = 1$ is also
shown in figure~\ref{fig:pdf_evolution_IB_e100}. For \pbat\, the asymptotic temperature distribution
will be $\varTheta=0$ whereas for \abpt\, will be $\varTheta=1$. 
\begin{figure}
\begin{center}
\includegraphics[width=0.68\linewidth]{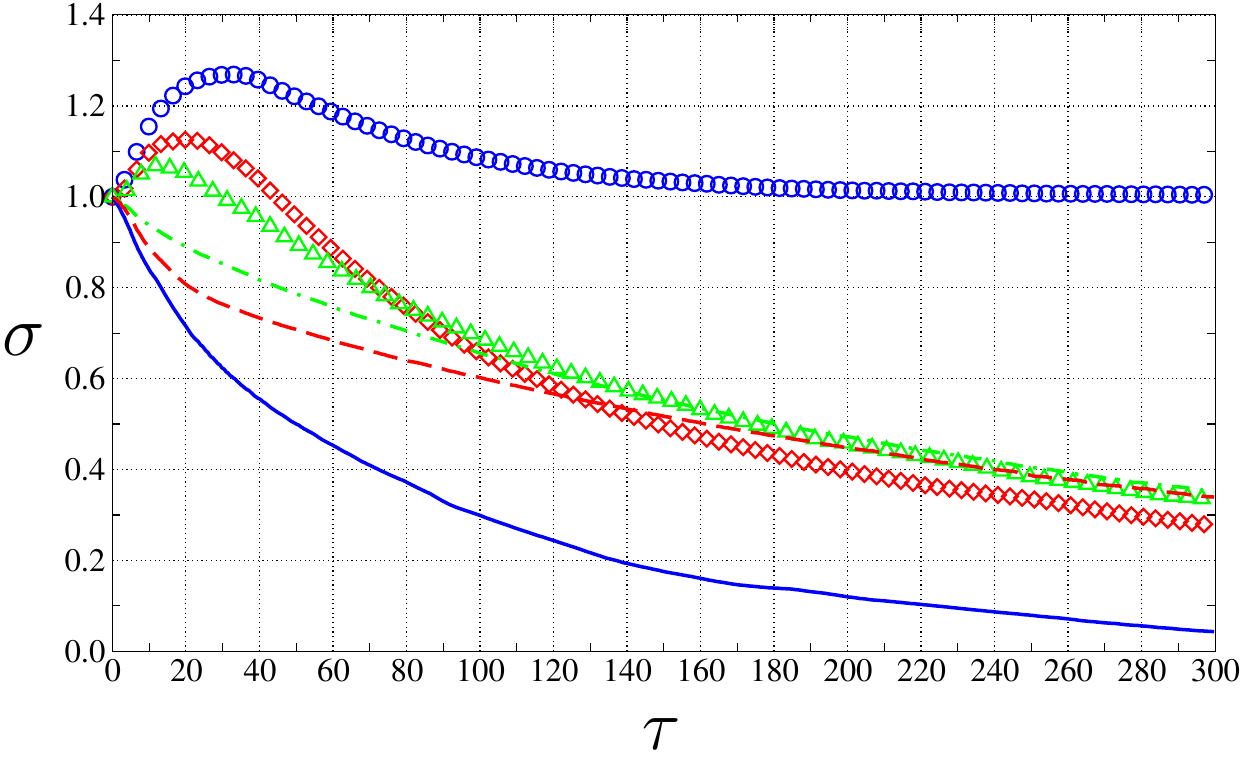}
\end{center}
     \caption{(Color online) Variance of temperature $\sigma$.
$\Delta$ (green online): \pbat and $\epsilon = 0.5$,  -- $\cdot$ -- $\cdot$ --  (green online) \abpt and $\epsilon = 0.5$;
$\circ$ (blue online): \pbpt and $\epsilon = 1$;  --- (blue online) \abat and $\epsilon = 1$; 
$\diamond$ (red online): \pbat and $\epsilon = 1$,  -- -- --  (red online) \abpt and $\epsilon = 1$.
}
\label{fig:variance}
\end{figure}

One of the main objectives of this study is the quantification of
mixing for several types of thermal boundary conditions on the horizontal walls. 
For \pbpt\, the flow mixes locally, but asymptotically, the flow returns to a state 
of linear stratification, therefore we can expect that this flow will not produce any
global mixing. The typical case of an initial value-decaying problem is
\abat\, where the final state of mixing is equal to the initial mean. 
For this case, it is custom to use the variance of temperature to quantify the mixing efficiency.
A uniform measure of mixing when sinks and sources are present is 
obtained by normalizing the variance by the value it would have in the absence of flow motion.
This `efficiency' measures how much mixing is increased by stirring, i.e.
if stirring decreases the variance compared to the value based purely on difussion
then the flow is mixed efficiently.

Figure~\ref{fig:variance} shows the variance for \pbat\, and \abpt\, at $\epsilon = 0.5$ and 1.
These norms are bounded below by a solid line (\abat) and above by circles (\pbpt) 
representing the best and worst mixing efficiencies (at $\epsilon = 1$) respectively.
For prescribed temperature at the bottom wall, the variance increases from its initial value
to a maximum and then decreases. The variance is larger in the higher $\epsilon$ case due to the more 
energectic spin-up that pushes more well-mixed cold fluid to the corner regions (compared to the
smaller $\epsilon$ value, generating a higher temperature contrast with the core.
The opposite effect is seen when the bottom wall is adiabatic, i.e.  the variance of temperature
is lower for $\epsilon = 1$ than for $\epsilon = 0.5$. This is also expected, since the
the amount of fluid and its temperature (carried to the corner region through the Ekman layer) 
is larger for the higher value of $\epsilon$.
The mixing features mentioned above seem to  agree with the flow 
similarities of \pbat\, with \pbpt\, and \abpt\, with \abat. 
The modified variance $\hat \sigma$ in figure~\ref{fig:hatvariance} demonstrates
how well the fluid mixes compared to the purely diffusive case for the same
conditions as figure~\ref{fig:variance}, where $\hat \sigma < 1$ corresponds to efficient mixing. 
Notice however, that after several tens of rotations the mixing generated by \pbat\, is unexpectedly smaller
than \abpt. Surprisingly, at $\tau=300$ the flow \abpt\, generates as much mixing as the pure diffusion case,
and excluding \abat, only \pbat\, at $\epsilon = 1$ generates $\hat \sigma < 1$ for $\tau > 300$.

\begin{figure}
\begin{center}
\includegraphics[width=0.88\linewidth]{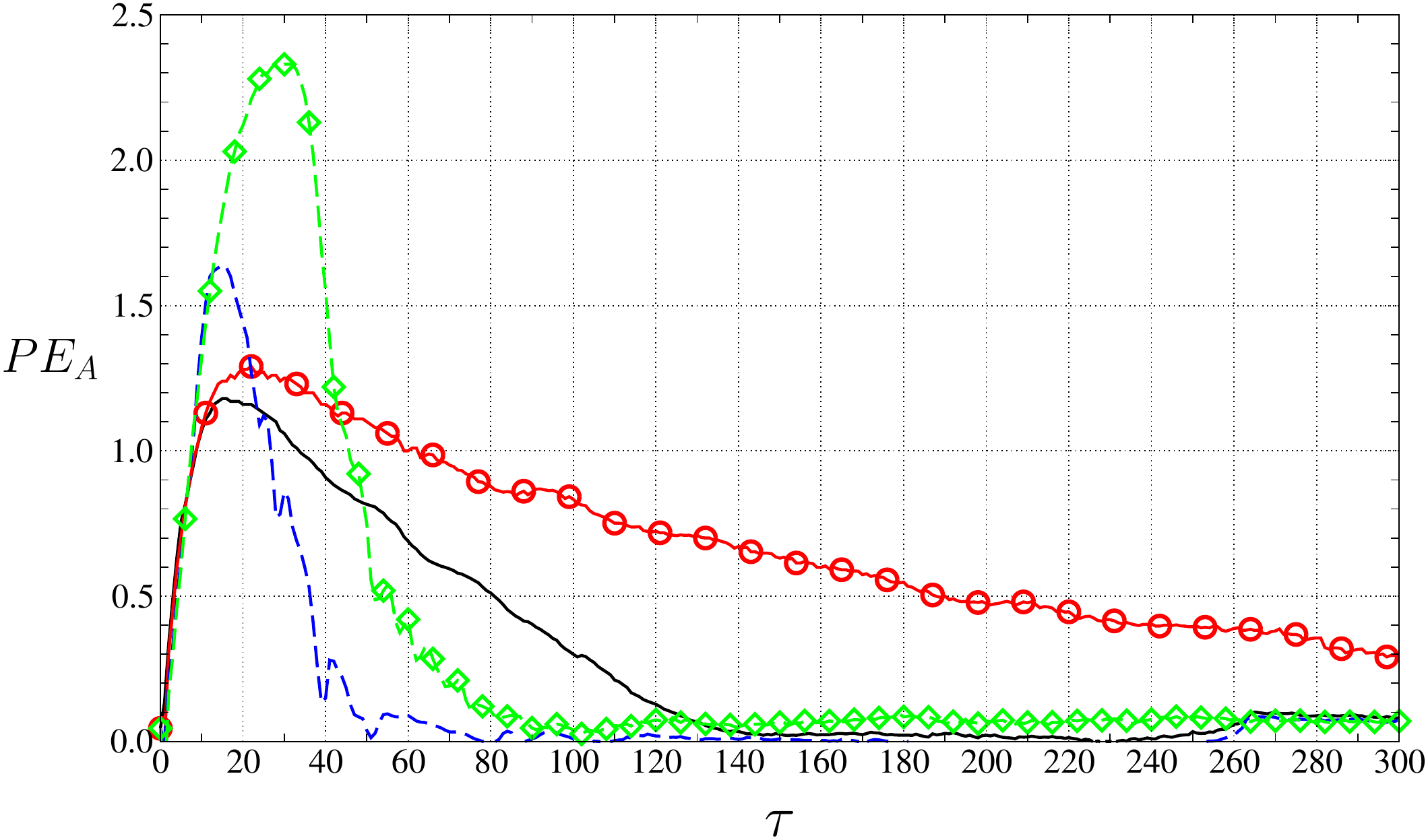}
\end{center}
     \caption{(Color online) Comparison of the potential energy available $PE_A$. 
              ----- (black), $\epsilon=0.5$ and \abpt; --$\circ$-- (red), $\epsilon=0.5$ and \pbat; 
              -- -- (blue), $\epsilon=1$ and \abpt; --$\diamond$-- (green), $\epsilon=1$ and \pbat. 
              }
\label{fig:PEA}
\end{figure}

Figure~\ref{fig:PEA} shows the time evolution of the potential energy available for mixing at
$\epsilon = \{0.5, 1\}$, for \pbat\, and \abpt. As expected, the available potential energy is
larger in the higher Rossby number due to more energetic stirring, and higher for \pbat\, than
for \abpt\, at the same $\epsilon$.  If a system has more potential energy available for mixing 
than another, then the system will mix better globally. This confirms our findings that
if \pbat\, has more potential energy available than \abpt\, (for the same $\epsilon$), then
asymptotically, \pbat\, will mix better than \abpt. The reason for \pbat\, to have more 
available potential energy than \abpt\, can  be explained as follows:
\begin{figure}
\begin{center}
\includegraphics[width=0.68\linewidth]{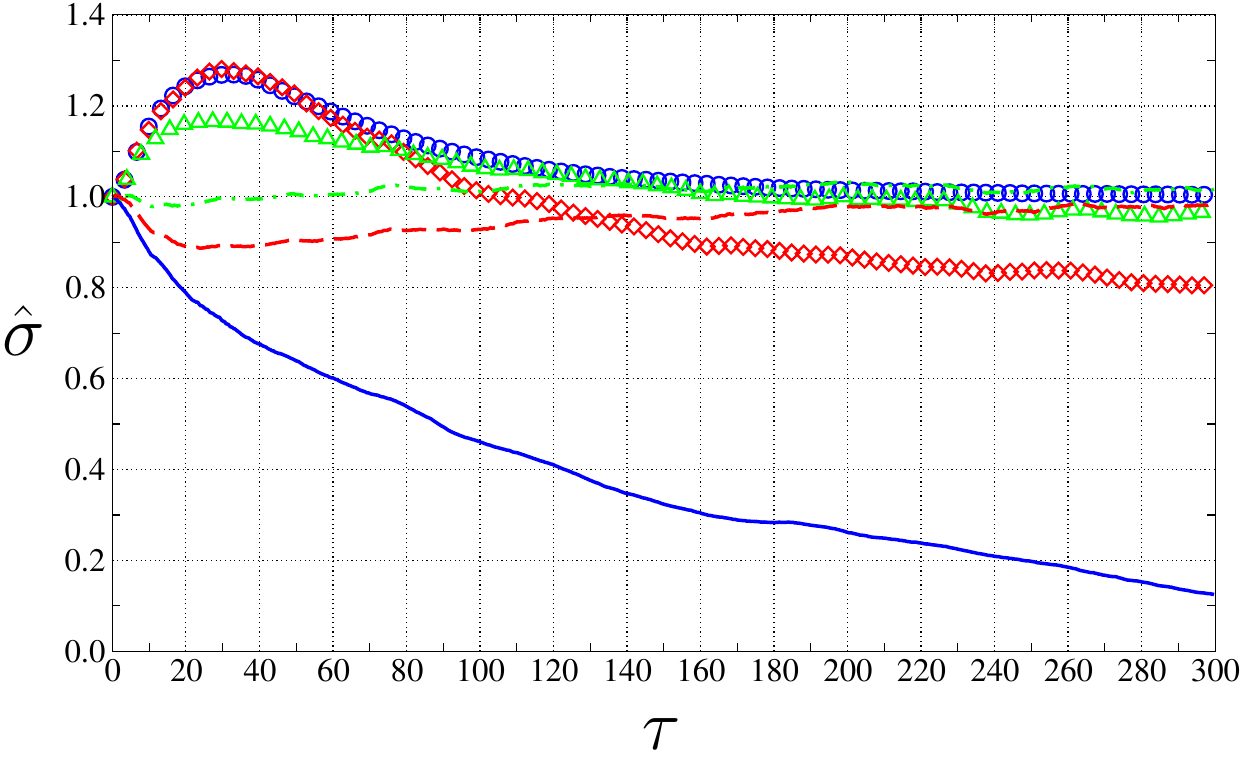}
\end{center}
     \caption{(Color online) Modified variance of temperature $\hat \sigma$.  
$\Delta$ (green online): \pbat\, and $\epsilon = 0.5$,  -- $\cdot$ -- $\cdot$ --  (green online) \abpt\, and $\epsilon = 0.5$;
$\circ$ (blue online): \pbpt\, and $\epsilon = 1$;  --- (blue online) \abat\, and $\epsilon = 1$; 
$\diamond$ (red online): \pbat\, and $\epsilon = 1$,  -- -- --  (red online) \abpt\, and $\epsilon = 1$.
}
\label{fig:hatvariance}
\end{figure}
for \pbat, the bottom wall is a sink of temperature, and during upwelling, 
the masses of fluid transported radially outwards through the Ekman layer cool down and accumulate 
at the corner regions (figure~\ref{fig:temperaturee100}). The corner regions are well mixed, but
only locally. These regions are separated from the core flow which remains in nearly solid body rotation.
The stirring caused by the upwelling also increases the temperature 
gradients and the potential energy that will be released when the Ekman transport shuts down.
The effect of prescribed bottom wall temperature also deteriorates the mixing but only during upwelling,
creating sharp gradients of temperatures among pockets of cold and relatively warm temperature. 
These gradients are higher for \pbat\, than for \abpt, because for \abpt, the bottom wall does
not cool down the fluid and thus the potential energy available for mixing for \pbat\, is
larger than for \abpt. Once the available potential energy is released, 
mixing will be generated by transforming the potential energy to kinetic energy. Thus the
higher the $PE_A$ the better the mixing.

\section{Conclusions}

In this paper we have studied numerically the mixing efficiency of spin-up 
stratified by temperature. Four different combinations of boundary conditions were 
considered at the bottom/top walls,  prescribed but fixed temperatures, adiabatic 
or a combination of these two.
The kinetic energy growth rate of the azimuthal disturbance was used to determine
when the baroclinic instability occurred. We found that the
spin-up with prescribed temperature at the bottom wall and adiabatic top wall 
was remarkably similar to the flow generated when the temperatures at the
horizontal walls were prescribed (\pbpt).
We focused on the quantifying the mixing using the variance
of temperature and a ratio of the variance to the value it would have
without stirring. When the temperatures are prescribed on the horizontal walls
the asymptotic state recovers its initial stratification, thus the effect
of spin-up worsens the global norm of mixing. 
When the walls were adiabatic, the flow achieved the highest efficiency of mixing.
The mixing efficiency for a flow with 
prescribed temperature on one wall and adiabatic on the other
yielded a mixing efficiency higher than \pbpt\, but lower than \abpt. 

Since the flow features for \abpt\, resembled those of \abat, and the
latter yielded the highest degree of mixing, we expected that the combination of
bottom adiabatic wall and prescribed temperature at the top would render better mixing than \pbat. 
This was true only for intermediate times, but asymptotically, \pbat always performed better than \abpt\,
(for the same $\epsilon$).  This was confirmed by evaluated the  potential energy available for mixing
for the two flows. During spin-up, the prescribed bottom-wall temperature cooled down the
fluid moving radially through the Ekman layer towards the corner regions, creating pockets of 
cold, but well-mixed fluid, keeping the potential energy
available for mixing at a higher level than that obtained through the
bottom adiabatic wall. This in turn created higher gradients of temperature, 
and therefore better mixing for large times.

There are many aspects of non-linear spin-up flows
that remain unexplored, and this study provides the
framework for further investigations.
One of them is how the mixing is affected by the
thermal diffusivity.
The thermal diffusion effect seems relevant for a period longer than the
Ekman spin-up time interval. If the thermal diffusion is large,
the azimuthal variations of temperature will decay quickly and the
the baroclinic term is likely to produce less vorticity.  But
whether or not a small thermal diffusion will render better mixing, 
is an open question.  Further investigation is also
needed on the effects of salt-stratification. These two effects
are currently being investigated.

\begin{acknowledgments}
  M. \, Baghdasarian is the recipient of a CEaS fellowship, from an NSF HRD-0932421 grant, for which we are
  grateful.  The comments of R\'emi Tailleux, Steve Wiggins and Jan-Luc Thiffeault  have greatly influenced the final 
  version of this paper and are
  very much appreciated.  We acknowledge the HPC and visualization resources provided by computing centre CASPUR
  {http://www.caspur.it} and the Ira A.\ Fulton A2C2 at Arizona State University.
\end{acknowledgments}

\end{document}